# Euler's variational approach to the elastica

Sylvio R. Bistafa
sbistafa@usp.br
Retired, Polytechnic School, University of São Paulo, Brazil

**Abstract**

The history of the elastica is examined through the works of various contributors, including those of Jacob and Daniel Bernoulli, since its first appearance in a 1690 contest on finding the profile of a hanging flexible cord. Emphasis will be given to Leonhard Euler's variational approach to the elastica, laid out in his landmark 1744 book on variational techniques. Euler's variational approach based on the concept of differential value is highlighted, including the derivation of the general equation for the elastica from the differential value of the first kind, from which nine shapes adopted by a flexed lamina under different end conditions are obtained. To show the potential of Euler's variational method, the development of the unequal curvature of elastic bands based on the differential value of the second kind is also examined. We also revisited some of Euler's examples of application, including the derivation of the Euler-Bernoulli equation for the bending of a beam from the Euler-Poisson equation, the pillar critical load before buckling, and the vibration of elastic laminas, including the derivation of the equations for the mode shapes and the corresponding natural frequencies. Finally, the pervasiveness of Euler's elastica solution found in various studies over the years as given on recent reviews by third parties is highlighted, which also includes its major role in the development of the theory of elliptic functions.

1. **Introduction**

On May 1690 Jacob Bernoulli (1655–1705) started a contest on finding the profile of a hanging flexible cord. By not specifying any condition which limits the problem to the nonelastic case, he challenges the mathematicians of the time to find the shape of a hanging elastic rope, on the belief that if one finds the curvature for the elastic case, then that of the nonelastic case can be obtained from it.

In the same year, Gottfried Leibniz (1646–1716) replied to Jacob Bernoulli's challenge saying that he would considered instead the shape of a thread hanging from its two points takes because of bending under its own weight, on the assumption that the thread, like a chain, keeps the same length and is neither stretched nor shortened as a normal thread would do. This assumption simplified the original challenge into finding the catenary curve. If the rope is elastic in all its parts, the problem turns into finding the *elastica*[a]. On the other hand, if the rope is as rigid as a chain, it turns into finding the catenaria[b].

From then on, the focus of the participating mathematicians was to find the curvature of the catenaria, all except Jacob Bernoulli who remained with his original problem. In the 1691 June issue of the *Acta Eruditorum*, three solutions were published: one by Jacob's brother Johann

---

[a] Latin for a thin strip of elastic material.

[b] The catenary is the equilibrium shape assumed by a chain suspended from two points. One speaks of a chain rather than any kind of string since a chain with very small links is fully flexible and unstretchable, as the idealized physical model assumes.

Bernoulli (1667–1748), along with two other solutions by Christian Huygens (1629–1695) and Leibniz (see Sepideh Alassi [1] for more details about these developments).

In the same issue of the *Acta Eruditorum*, Jacob Bernoulli posed the precise problem of the elastica (see Raph Levien [2]):

> "Assuming a lamina $AB$ of uniform thickness and width and negligible weight of its own, supported on its lower perimeter at $A$, and with a weight hung from its top at $B$, the force from the weight along the line $BC$ sufficient to bend the lamina perpendicular, the curve of the lamina follows this nature: The rectangle formed by the tangent between the axis and its own tangent is a constant area."

This poses one specific instance of the general elastica problem, now generally known as the rectangular elastica, because the force applied to one end of the curve bends it to a right angle with the other end held fixed (Fig.1).

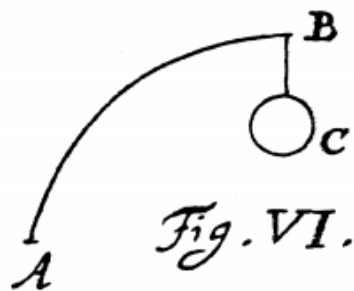

Figure 1: In 1691 Jacob Bernoulli poses the elastica problem.

By 1692, Jacob Bernoulli working on a geometric construction of the curvature gives the differential equation of the curve $AB$ in a readily computable form; $y$ can be obtained as the integral of a straightforward function of $x$ from

$$dy = \frac{x^2 dx}{\sqrt{a^4 - x^4}},$$

saying that he would be showing it in due time.

Indeed, two years later, in 1694, Jacob Bernoulli published in *Acta Eruditorum* the geometric construction for the elastica, confirming the above result.

As Alassi [1] points out, at the time, a solution to a problem would only be accepted if a geometric construction was provided. Descartes introduced this problem-solving technique in 1637 in *La Géométrie*, which was then widely used by the mathematicians. This technique considers that the intersections between algebraic curves were enough to solve all kinds of geometrical problems. However, mathematicians found that other geometric curves, such as a hyperbola, would make constructions easier. For instance, Leibniz used a logarithmic curve in his geometric construction of the catenaria. Mechanical devices such as levers and pulleys were also used, and Jacob Bernoulli used a pulley in his geometric solution of the velaria and a lever in his geometric construction of the elastica.

The geometric construction by quadrature[c] was explained by Alassi [1] on a simplified version of Jacob Bernoulli's geometric construction of the elastica as follows. An elastic band is fixed at one end and a weight is suspended from its free end $O$ (Fig. 2), and the shape that this elastic band takes, the elastica, is sought. The main difficult in this procedure is to find the auxiliary curve. However, in physical problems such as the elastica, the auxiliary curve can be found by identifying the physical characteristics of the system. For example, the elasticity must be defined geometrically as a curve describing the relation between stress and strain. Once this curve is found, the rest of the construction follows.

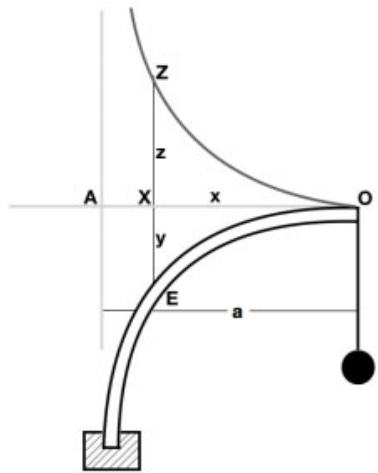

Figure 2: Simplified illustration of Jacob Bernoulli's construction of the elastica (Bos, 1986, as cited in Alassi [1])

In a very complicated development, and with the help of the auxiliary curve $OZ$, Jacob Bernoulli eventually shows that the ordinate $y$ on the elastic curve is found from the integral (see Lawrence D'Antonio [3] for more details about this development)

$$y = \int_0^x \frac{x^2 dx}{\sqrt{a^4 - x^4}}.\text{[d]}$$

Jacob Bernoulli conjectured that this integral could not be expressed in terms of 'known' functions, sin, exp, sin$^{-1}$, and succeeded in finding a series solution which is given by[e]

---

[c] In the 17th century, finding the area of a figure was often called quadrature, or squaring, which is equivalent to the integral.

[d] This integral is known as an elliptic integral. The study of elliptical integrals can be said to start in 1655 when Wallis began to study the arc length of an ellipse. In fact, he considered the arc lengths of various cycloids and related these arc lengths to that of the ellipse. Both Wallis and Newton published an infinite series expansion for the arc length of the ellipse. More information on elliptic integrals and elliptic functions cab be found, for example, at
http://www.mhtlab.uwaterloo.ca/courses/me755/web_chap3.pdf. Accessed Nov 11, 2022.

[e] Jacob Bernoulli demonstrates this in the *Meditationes CLXXV*:
https://beol.dasch.swiss/transcription/http:%2F%2Frdfh.ch%2F0801%2FsobTv3-LT4m_Q5MBtFA8Xw
See also Alassi [1].

$$y = \int_0^x \frac{x^2 dx}{\sqrt{a^4 - x^4}} = \frac{x^3}{3a^2} + \frac{x^7}{2 \cdot 7 \cdot a^6} + \frac{1 \cdot 3 \cdot x^{11}}{2 \cdot 4 \cdot 11 \cdot a^{10}} + \frac{1 \cdot 3 \cdot 5 \cdot x^{15}}{2 \cdot 4 \cdot 6 \cdot 15 \cdot a^{14}} + \cdots$$

Very shortly after Jacob Bernoulli's publication, Huygens published a short note in the same forum illustrating several of the possible shapes the elastica might take on and pointing out that Bernoulli's quadrature only expressed the rectangular elastica. The shapes are shown from left to right in order of increasing force at the endpoints, and shape A is clearly the rectangular elastica (Fig. 3).

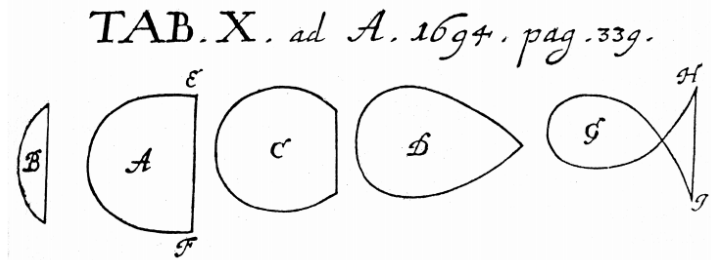

Figure 3: Huygens's 1694 several possible shapes where shape A is Jacob Bernoulli's solution.

Jacob Bernoulli acknowledged this criticism and indicated that his technique could be extended to handle these other cases by using a non-zero constant for the integration. Nonetheless, more than 40 years had passed for a definitive solution to the elastica problem by Euler's analysis of 1744.

## 2. Daniel Bernoulli and Leonhard Euler elastica

In 1724, an anonymous author restarted Jacob Bernoulli's challenge of 1690 by proposing the unification of the catenaria and the elastica. This challenge was left unattended for years until Daniel Bernoulli (1700–1782) and Leonhard Euler (1707–1783), working together in a friendly competition, published in the same 1732 February issue of *Commentarii Academiae Scientiarum Petropolitanae* [4, 5] reciprocal contributions to this challenge.

Daniel Bernoulli in his 1732 treatise *Methodus universalis determinandae curvatura fili* ... [4] considered an elastic band fixed at one end, which bends partially under its own weight and partly through a suspended load at the free end. The moment of the suspended weight is $Px$, where $x$ is the distance from the free end to an arbitrary point of the band, and the moment of the weight of the band is given by $\sigma \int s dx$, where $\sigma$ is the weight/unit of length of the band (supposedly uniform). In equilibrium, the total moment of these external forces should be balanced by the moment of the internal forces of the elastic band $\mathcal{M}$, which Daniel Bernoulli postulates as being related to the radius of curvature $r$ at an arbitrary point of the band by the formula

$$\mathcal{M} = \frac{\beta}{r},$$

where $\beta$ is the modulus of bending[f].

As we shall see next, Euler adopted essentially the same expression in his derivation of the elastica as given in his 1732 treatise [5], and this result became the first explicit definition of the fundamental law of the elastica[g]. This result is known today as the *Euler-Bernoulli formula* for the bending of a beam, recording the achievements of Euler and Daniel Bernoulli. In fact, Jacob Bernoulli had implied in his extensive work with the elastica that the load is inversely proportional to the radius of curvature $r$, however, he never stated the law of the elastica in that way (see Alassi [1] for more details).

---

[f] Nowadays, this formula is written as $\mathcal{M} = \frac{EI}{r}$, where $E$ is the Young's module, and $I$ is the second moment of area or area moment of inertia. The product $EI$ is known as the flexural rigidity.

Eventually, Daniel Bernoulli in his 1732 treatise [4] obtains $\mathcal{M} = \sigma \int s dx + Px$, and finally,

$$\sigma \int s dx + Px = \frac{\beta}{r}.$$

Daniel Bernoulli left the study of the curvature to Euler; he stated: "… whatever can be thought out about the kind of curve present, which our own illustrious Euler correctly observed, who himself thus had proposed this problem to be solved, so that nothing can be added."

Euler is known to be the one who unified the catenary problem and the elastica problem in his 1732 treatise *E8 -- Solutio problematis de invenienda curva, quam format lamina utcunque elastica* ... [5], stating at the Introduction "… The most famous Jacob Bernoulli was the first, and later many others, to assign a curve to curved elastic strips, which is known by the name of the elastic curve (the elastica), where it is understood that their solutions are only applicable to elastic strips without weight. Although the curvature of a heavy elastic strips may take place in the realm of nature, yet, as far as I know, no one has yet determined it …" Euler then reveals the scope of his work "… Indeed the first investigation extends this far, that the curve formed by a heavy laminar shape, with one end fixed and having some force applied to the other end, can be found. However, at the start of the work I consider a more general lamina of arbitrary elasticity, and having some hanging weights attached …"

Like Jacob Bernoulli's geometric construction of the elastica, Euler also uses an auxiliary curve in his developments. He begins by considering $AM$ as a curved lamina (Fig. 4), with $A$ as the origin of a left-going x-axis, and with forces applied at all points $N$, given by the *auxiliary curve* $BG$ as $g(x)$ which represents a downward force/unit of length, due perhaps to the weight of the lamina, such that the area $AQHB$ is found by $h = \int g(x')dx'$. Then the sum of all the moments of all the forces to the curve $ANM$ turning about $M$ shall be as the area $APT$ …"

The moment at $M$ of all the forces up to $P$, whose coordinate with respect to some origin is $x$, is given by $M(x) = \int_x^X x' g(x')dx'$, where the coordinate of point $A$ is $X$. The integration is performed by parts as $M(x) = [xh(x)]_x^X - \int_x^X h(x')dx' = \int_X^x h(x')dx'$, since either $x = 0$ or $h(X) = 0$. This result is represented by the area $APT$ above the curve $AVT$.

By calling a general point $x$ on this curve by $P = P(x)$, hence the sum of all the moments at $M$ will be equal to $\int P dx$.

Next, the radius of curvature at the point $M$ is put equal to $r$; and the angle that the two elements constitute vary inversely as $r$. The elastic strength at $M$ is designated by the letter $v$; the strength producing this angle varies as $\frac{v}{r}$ (hyp.). As shown above, the sum of the moments of the distributed vertical forces is equal to $\int P dx$ and, similarly, the sum of the moments of the distributed horizontal forces $\int Qdx$. By calling $E$ and $F$ localized vertical and horizontal forces distanced $x$ and $y$ from $M$, respectively, thus, the total strength of the moments acting at $M$ is equal to $Ex + Fy + \int Pdx + \int Qdy$. Since $\frac{v}{r}$ should be proportional to that moment, the equation is obtained: $\frac{Av}{r} = Ex + Fy + \int Pdx + \int Qdy$, where $A$ is a constant.

---

[g] In Daniel Bernoulli's derivation "… he does no more than restate it … for Daniel Bernoulli, as with most principles he considered true, it seems to be self-evident and scarcely worthy a comment …" *In The Rational Mechanics of Flexible or Elastic Bodies*, C. Truesdell, 1960, p.147.

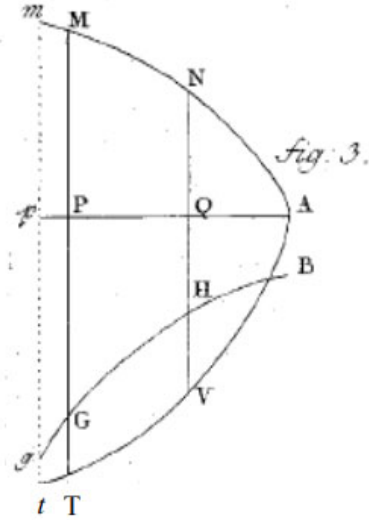

Figure 4: A curved lamina $AM$ and the auxiliary curve $BG$. Source: Euler [5], fig. 3.

Then, after a heavy mathematical manipulation of this result, Euler develops the equation that gives all the possible curves which perfectly flexible bodies ($v = 0$) can form by being displaced in some manner. However, because of the forbidding appearance of the expressions developed by Euler, we will refrain to show them here (for more details, see Ian Bruce's translation of Euler's 1732 treatise [5]).

Next, Euler poses several problems, in which he develops solutions for catenaries, awnings, and sails. It is only at the end of the 1732 treatise that Euler considers finding the curve $BMA$ that an elastic strip fixed at $B$ and with a free end at $A$ forms under its own weight, from the simplified form of the general equation

$$\frac{Av}{r} = \int Pdx,$$

where $A$ is a constant, $v$ is the elastic strength, and $r$ is the radius of curvature at a given point $M$ of the strip.

For a wire with the same weight/unit length everywhere, $dP$ is constant and equal to $ads$ ($P = as$), and since the elasticity is everywhere the same, we may put $v = b$; and the equation becomes

$$\frac{Ab}{r} = \int asdx,$$

which, by considering that $\frac{1}{r} = \frac{d^2x}{dsdy}$, $dy = ds\sqrt{1-p^2}$, and getting rid of the superfluous constants $a$ and $b$ results in

$$sp^2(1-p^2)^{\frac{3}{2}}ds^2 = Apdp^2 + A(1-p^2)d^2p.$$

Regrettably, Euler considers that even these adaptations, do not allow the construction of the curve.

## 3. Euler's variational analysis of the elastica

The ideas behind the use of variational techniques to solve the elastica problem appear in a series of letters between Daniel Bernoulli and Euler (see Lawrence D'Antonio [3] for more details about this epistolary exchange). The first mention was in a 7 March 1739 letter, suggesting the "isoperimetric method" (an early name for the calculus of variations) for the elastica problem[h]. However, the first clear mathematical statement of the elastica, as a variational problem in terms of the stored energy, appears in a letter of October 1742 where Daniel Bernoulli conjectured that the minimum of the *strain potential energy* (in modern terminology) of a curved elastic lamina expressed as $\int \frac{ds}{R^2}$, assuming the element of the arclength $ds$ is constant and indicating the radius of curvature by $R$, must result in the elastica. He admitted that he had tried to find this minimum point himself, but his investigation had resulted in a differential equation of the fourth order, which he could not reduce to a lower-order one (see Alassi [1]). Daniel Bernoulli then invited to Euler to try his skills on the minimization: "… There is nobody as perfect as you [Euler] for easily solving the problem of minimizing $\int \frac{ds}{R^2}$ using the isoperimetric method."[i] This invitation resulted in Euler's major research on elasticity, which was published in 1744 as *Additamentum I*, *De Curvis Elasticis* added as an appendix to his famous treatise *Methodus inveniendi curvas lineas* [6], a treatise on the calculus of variations.

*Methodus inveniendi curvas lineas* is considered the fundamental work in the calculus of variations and, as such, has been examined by several modern investigators. Nonetheless, a complete annotated translation by Ian Bruce has only recently become available over the Internet[j]. This translation is introduced with a brief summary of the six chapters contained in the treatise, plus two appendices, in which Appendix A (*Additamentum I*) deals with curves associated with elastic laminas, which was later added to the main work by Euler. For convenience, Ian Bruce has subdivided the translation of *Additamentum I* in two appendices 1A and 1B. This is how he summarizes the content of the former: "… Euler sets out to show that his method of finding maxima or minima curves associated with generalized functions in the form of integrals, can be applied to finding the shape of loaded laminas or ribbons, as had then recently been established in a straightforward method from mechanics by Daniel Bernoulli, following on the earlier work of his uncle, Jacob Bernoulli. Most of the first part of this appendix, so subdivided for convenience, is given over to finding the nine classes or kinds of shapes adopted by a flexed lamina under different end conditions. An English translation exists already in Isis (1933) by Oldfather et al., of which I have just become aware, and have not referred to here …" Ian Bruce completes his contribution with two additional translations of Euler's later overview of his main treatise on the calculus of variations: E296 "The elements of the calculus of variations", and E297 "The analytical explanation of the method of maxima and minima."

---

[h] According to Levien [2]: "… Many founding problems in the calculus of variations concerned finding curves of fixed length (hence isoperimetric), minimizing or maximizing some quantity such as area enclosed. Usually, additional constraints are imposed to make the problem more challenging, but, even in the unconstrained case, though the answer (a circle) was known as early as Pappus of Alexandria around 300 A.D, rigorous proof was a long time coming."

[i] Levien [2] indicates that Truesdell also points out that this formulation wasn't entirely novel; Daniel Bernoulli and Euler had corresponded in 1738 about the more general problem of minimizing $\int R^m ds$, and they seemed to be aware that the special case $m = -2$ corresponded to the elastica.

[j] https://17centurymaths.com/contents/Euler'smaxmin.htm

One of the modern analysis of *Methodus inveniendi curvas lineas* is the article by Craig G. Fraser, "The Origins of Euler's Variational Calculus" [7], who appraises the work as "… Euler succeeded in formulating the variational problem in a general way, in identifying standard equational forms of solution and in providing a systematic technique to derive them. His work included a classification of the major types of problems and was illustrated by a wide range of examples …"

An explanation of Euler's derivation of the general equation for elastic laminas that we have been able to identify is the one given by Clifford Truesdell [8, pp. 203–204]. However, despite being a fundamental contribution to understand Euler's formulation, we feel that more could be said regarding the introductory part of the application of the variational formulation to the elastica given by Euler in the derivation of the general equation.

Indeed, the method employed by Euler was laid out in Chapter IV of the *Methodus inveniendi*, which begins with the proposition: *To find the equation between the two variables $x$ and $y$, thus so that for a given value of $x$, for example on putting $x = a$, the formula $\int Z dx$ shall obtain a maximum or minimum value, with $Z$ being a function of $x, y, p, q, r$ etc. either determinate or indeterminate*. As usual: $p = \frac{dy}{dx}, p = \frac{dp}{dx}, r = \frac{dq}{dx}, s = \frac{dr}{dx}$ etc.

Next, Euler gives what he calls the *differential values* of the formula $\int Z dx$ for various kinds of the function $Z$, which may correspond always to the magnitude of the variable $x$, considered to be $x = a$. The d*ifferential value* is a key concept in Euler's method of calculus of variations, which is defined as: "… The differential value of a given maxima or minima corresponding to a formula is the difference between the values, which this formula may maintain both on the curve itself, as well as on the same changed by an infinitely small amount." In Chapter IV, five kinds of differential values of increasing complexity are given by Euler, from where the *differential value of the first kind* was taken in the development of the general equation for the elastica.

Suppose that the curve $az$ in Fig. 5 is such that the integral $\int Z dx$ is a maximum or minimum. A comparison curve $amvoz$ is then obtained by increasing the ordinate $Nn$ by an infinitely small quantity $nv$.

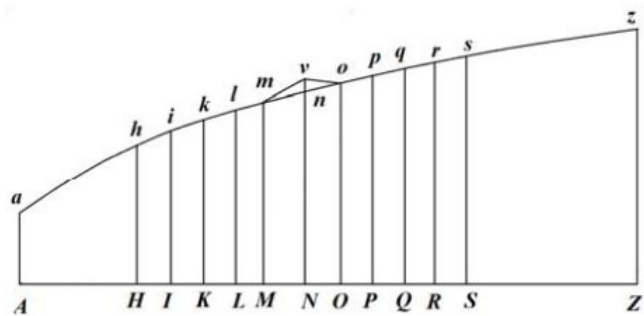

Figure 5: Geometrical elements used by Euler to derive the expressions for the differential values associated with integrals of the form $\int Z dx$. (Source: Figure 4 of *Methodus inveniendi*)

Euler writes the total differential of $Z$ in the following form

$$dZ = Mdx + Ndy + Pdp + Qdq + Rdr + etc,$$

where $M = \frac{\partial Z}{\partial x}$, $N = \frac{\partial Z}{\partial y}$, $P = \frac{\partial Z}{\partial p}$, $Q = \frac{\partial Z}{\partial q}$, $R = \frac{\partial Z}{\partial r}$, etc.

He then derives the *differential value of the first kind* for the general formula $\int Z dx$ as

$$nv \cdot dx \left( N - \frac{dP}{dx} + \frac{d^2 Q}{dx^2} - \frac{d^3 R}{dx^3} + etc \right),$$

which is a result that Euler applies in several examples throughout his treatise.

Right at the beginning of *Additamentum I* of *Methodus inveniendi*, after having mentioned that Daniel Bernoulli had indicated to him that the general force (potential energy), which is present in a curved elastic lamina, can be included in a single expression which he calls the *force potential*, Euler sets out his objective "… *so that amongst all the curves of the same length, which not only may pass through the points A and B, but also may be tangents at these points with right lines given, that may be defined in which the value of this expression $\int \frac{ds}{R^2}$ shall be a minimum.*"

Here, $ds$ is the element of the curve, given by $ds = \sqrt{1 + p^2}$, and $R$ is the radius of curvature, which is given by the well-known formula $R = \frac{(1+p^2)^{3/2}}{q}$, where $p = \frac{dy}{dx}$ and $q = \frac{dp}{dx}$. From these, two integrals of the general form $\int Z dx$ are introduced: $\int dx \sqrt{1 + p^2}$ $(Z = \sqrt{1 + p^2})$, which comes from the $\int ds$, and represents the isoperimetric condition, and $\int \frac{q^2 dx}{(1+p^2)^{5/2}}$ $(Z = \frac{q^2}{(1+p^2)^{5/2}})$, which comes from $\int \frac{ds}{R^2}$, and this is the integral that must be a minimum.

From the general expression for $dZ = Mdx + Ndy + Pdp + Qdq$, the *differential value* associated with the integral $\int dx \sqrt{1 + p^2}$ is $nv \cdot dx \left(-\frac{dP}{dx}\right) = -nv\, dP = -nv\, d.\frac{p}{\sqrt{1+p^2}}$. By its turn, the *differential value* associated with the integral $\int \frac{q^2 dx}{(1+p^2)^{5/2}}$ is $nv \cdot dx \left(-\frac{dP}{dx} + \frac{d^2 Q}{dx^2}\right) = nv \left(-dP + \frac{dQ}{dx}\right)$, where $P = \frac{-5pq^2}{(1+p^2)^{7/2}}$ and $Q = \frac{2q}{(1+p^2)^{5/2}}$.

On account that these two differential values are associated with the same curve, they should be equal, and the equation for the sought curve is then

$$\alpha\, d.\frac{p}{\sqrt{1 + p^2}} = dP - \frac{dQ}{dx},$$

where $\alpha$ was introduced as a scaling factor, which was later identified as a Lagrange multiplier [2]. This equation is essentially the starting point of Truesdell explanation of Euler's derivation of the general equation for the elastica [8, pp. 203–204].

This is how Daniel Bernoulli reacts to the solution of the fourth-order differential equation obtained by him earlier for the elastica in the October 1742 letter to Euler: "… *given my assumption that the potential energy of the elastic lamina must be minimal, as I've mentioned to you before. In this way I get a 4th order differential equation, which I have not been able to reduce enough to show a regular equation for the general elastica.*"

Indeed, the integration of the equation was accomplished by Euler with a long and hard to follow *ad hoc* procedure (for more details see Ian Bruce's annotated translation of *Additamentum I*), to which Truesdell [8] highlights the main results, and eventually presents Euler's final solution as

$$dy = \frac{(a^2 - c^2 + x^2)dx}{\sqrt{(c^2 - x^2)(2a^2 - c^2 + x^2)}}.$$

It is now known that the integration of this equation may be expressed by elliptic functions[k]. However, Euler follows a simple and direct analysis for different relations between $a$ and $c$, which in some cases the integration was accomplished by series expansions or that result in simple integrable formulas.

Euler then seeks in the section *Enumeratio curvarum elasticarum* to determine the "infinite variety of these elastic curves." The above equation is the basis for Euler's graphical study of the elastic curve. With the aid of asymptotic behaviors and geometric relations Euler finds nine different curves, which were classified by Levien [2] in terms of the parameter $\lambda = \frac{a^2}{c^2}$ and that are enumerated in the table below:

| Euler's species # | Euler's Figure | Euler's parameters | $\lambda$ | comments |
|---|---|---|---|---|
| 1 | | $c = 0$ | $\lambda = \infty$ | straight line |
| 2 | 6 | $0 < c < a$ | $0.5 < \lambda$ | |
| 3 | | $c = a$ | $\lambda = 0.5$ | rectangular elastica |
| 4 | 7 | $a < c < a\sqrt{1.651868}$ | $.302688 < \lambda < .5$ | |
| 5 | 8 | $c = a\sqrt{1.651868}$ | $\lambda = .302688$ | lemnoid |
| 6 | 9 | $a\sqrt{1.651868} < c < a\sqrt{2}$ | $.25 < \lambda < .302688$ | |
| 7 | 10 | $c = a\sqrt{2}$ | $\lambda = .25$ | syntractrix |
| 8 | 11 | $a\sqrt{2} < c$ | $0 < \lambda < .25$ | |
| 9 | | $a = 0$ | $\lambda = 0$ | circle |

Euler includes figures for six of the nine cases. Of the three remaining, species #1 is a degenerate straight line, and species #9 is a circle. Species #3, the rectangular elastica, is of special interest, so it is rather disappointing that Euler did not include a figure for it (see Levien [2], D'Antonio [3], Truesdell [8, pp. 205–210], and Fraser [9] for further discussions about these curves). Fraser [9] has additionally compared the classificatory scheme that Euler adopts for his transcendental general equation with the classificatory scheme that Newton employs for algebraic curves.

---

[k] Elliptic integrals can be viewed as generalizations of the inverse trigonometric functions and provide solutions to a wider class of problems. For instance, while the arc length of a circle is given as a simple function of the parameter, computing the arc length of an ellipse requires an elliptic integral. Similarly, the position of a pendulum is given by a trigonometric function as a function of time for small angle oscillations, but the full solution for arbitrarily large displacements requires the use of elliptic integrals. Many other problems in electromagnetism and gravitation are solved by elliptic integrals. From Wolfram Math World https://mathworld.wolfram.com/EllipticIntegral.html. Accessed Nov 11, 2022.

Euler's figures (Source: L. Euler [6])

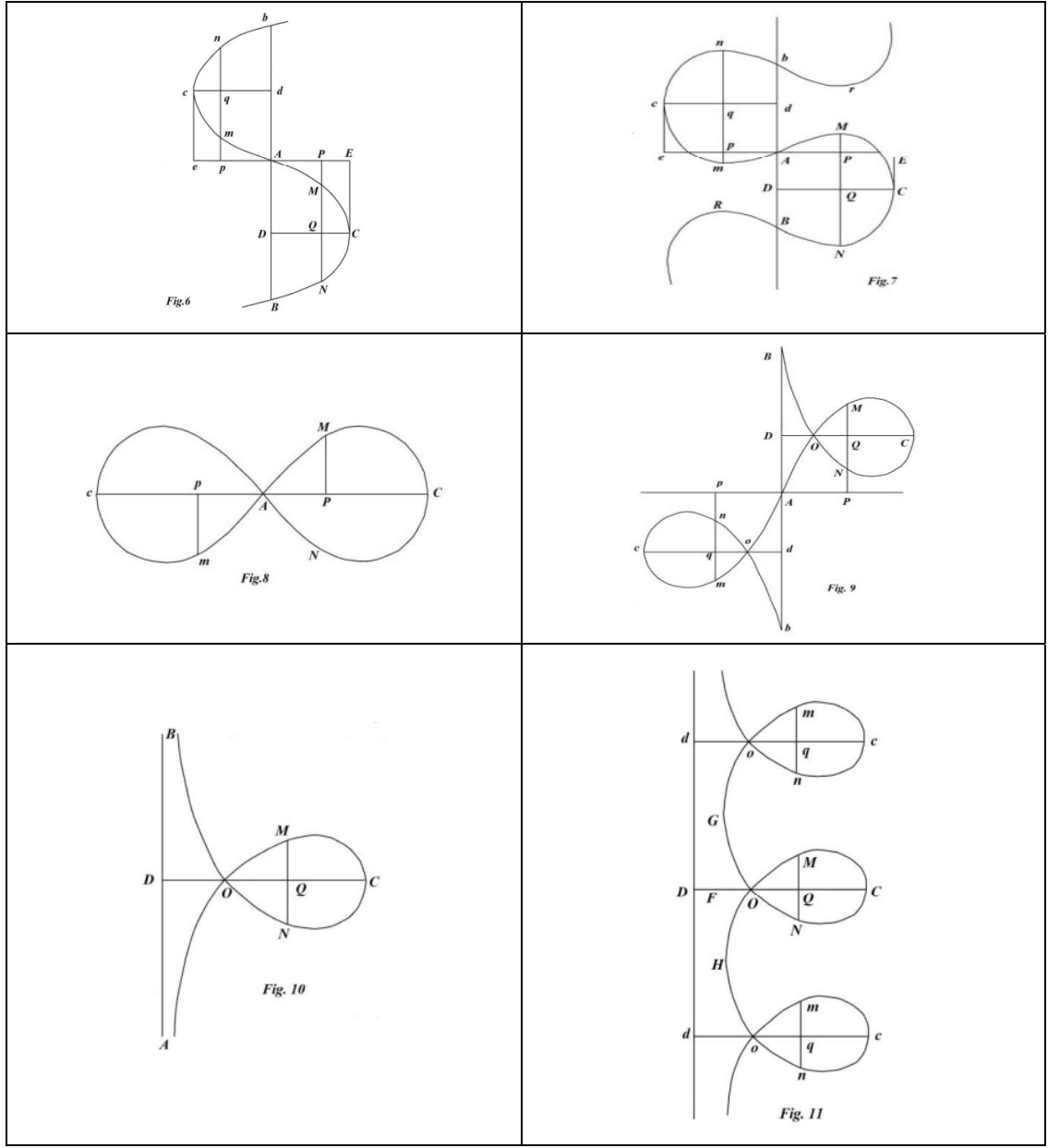

In the second part of *Additamentum I*, from the degenerate straight line (species #1), Euler obtains the well-known formula for the critical load $P_{cr}$ for a straight vertical column, given by

$$P_{cr} = \frac{\pi^2 EI}{l^2},$$

where $l$ is the length of the column, and, as before, $EI$ is the flexural rigidity.

The column will remain straight for loads less than the critical load. The critical load is the greatest load that will not cause lateral deflection (buckling); for loads greater than the critical load, the column will deflect laterally (further discussions on the so-called Column Problem can be found in Refs. [3, 8, 9]).

**The Euler-Poisson equation**

Although this topic was a later development and not directly considered by Euler in his treatise, the time is ripe to show an application of the Euler-Poisson to the elastica. As already considered earlier, the elastica treated as a variational problem becomes a problem of finding a minimum for

$$\int \frac{ds}{R^2} = \int \frac{q^2 dx}{(1+p^2)^{\frac{5}{2}}}.$$

Since $p = \frac{dy}{dx}$ and $q = \frac{dp}{dx}$, it is seen that this equation is written in terms of the first and second derivatives of $y(x)$, and so the simple Euler-Lagrange equation does not suffice. In this case, we may use the Euler-Poisson equation, which when the functional contains second order derivatives may be written as[i]

$$\frac{\partial Z}{\partial y} - \frac{d}{dx}\left(\frac{\partial Z}{\partial p}\right) + \frac{d^2}{dx^2}\left(\frac{\partial Z}{\partial q}\right) = 0.$$

As an example, let us find the shape of a cantilever beam under a uniform distributed load, by applying the principal of minimum (total) potential energy. This principle states that the total potential energy for such a system is a minimum with respect to all small displacements that satisfy the given boundary conditions.

In this example, the external force is the familiar gravitational force. The internal forces arise from strains within the beam. The total potential energy has two components: elastic strain energy and gravitational potential energy.

As before, the elastic strain energy comes from $\int \frac{ds}{R^2}$, which for a beam with bending stiffness $EI$ may be obtained from $\int \frac{EI}{2} \frac{q^2 dx}{(1+p^2)^{5/2}}$. As far as the gravitational potential energy is concerned, it must be first realized that when the beam is deflected, there is a loss in potential energy due to the external, uniform distributed load $\mu$, which may be written as $-\int \mu y \, ds = -\int \mu y (1 + p^2)^{1/2} dx$. Therefore, the beam total potential energy will be given by

$$\int \left[\frac{EI}{2} \frac{q^2}{(1+p^2)^{\frac{5}{2}}} - \mu y (1+p^2)^{\frac{1}{2}}\right] dx.$$

By assuming that the beam deflection is small, we can neglect second degree terms in $p$ and write the total potential energy as

$$\int \left(\frac{EI}{2} q^2 - \mu y\right) dx.$$

Then, the Euler-Poisson equation for this integral, with the functional $Z$ given by $\frac{EI}{2} q^2 - \mu y$, reduces to

$$EI \frac{d^4 y}{dx^4} = \mu.$$

This equation is known as the *Euler-Bernoulli* equation and describes the deflection of a uniform, static beam, under a distributed load. It is used widely in engineering practice, and tabulated expressions for the deflection for common beam configurations can be found in engineering handbooks.

---

[i] For a derivation of this result see, for instance, Kot, M. A First Course in the Calculus of Variations. Student Mathematical Library, Volume 72. American Mathematical Society, Providence, Rhode Island, 2014, Chapter 4, p 69.

In the second part of *Additamentum I*, Euler deals additionally with the following topics: 'The determination of the elastic curve by experiment', 'The unequal curvature of elastic laminas', 'Concerning the curvature of elastic laminas not naturally straight', 'The curvature of elastic laminas at individual points arising from whatever forces acting', 'The curvature of elastic laminas arising from their own weight', 'Concerning the oscillatory motion of elastic laminas', and the following sub cases of the latter: 'Oscillations of an elastic lamina with one end fixed to a wall', 'Oscillations of free elastic laminas', 'Oscillations of an elastic lamina fixed at both ends', and 'Oscillations of a lamina with each end fixed to a wall'. The difference between ends fixed and ends fixed to a wall, is that in the former, there are no resisting moments at the fixed points – the ends are simply pinned –, whereas in the former, there are resisting moments at the fixed points – the ends are clamped. These represent Euler's *tour de force* on trying to exhaust applications of his elastica analysis.

Here, we will only examine further the topic 'The unequal curvature of elastic laminas', to show the potential of Euler's variational approach. Further discussion of this and other cases listed above can be found in Truesdell [8].

**The unequal curvature of elastic laminas**

The problem here is to find the curvature of elastic laminas with non-uniform flexural rigidity. This problem then becomes to find a minimum for $\int \frac{Sds}{R^2}$, where $S$ is the flexural rigidity which is now allowed to vary along the lamina. In this case, Euler applies the procedures laid down in Chapter 3 of the *Methodus inveniendi*.

This development begins in § 41, where Euler writes $dS = Tds$, and as usual, $dy = pdx$, $dp = qdx$, and among all the curves, in which $\int dx\sqrt{1+p^2}$ is of the same magnitude, that must be determined, in which $\int \frac{Sq^2}{(1+p^2)^{5/2}}$ must be a minimum, and recalling that $R = \frac{(1+p^2)^{3/2}}{q}$.

As found earlier, the *differential value* associated with the integral $\int dx\sqrt{1+p^2}$ is

$$d.\frac{p}{\sqrt{1+p^2}}. \qquad (I)$$

However, it should be realized that whereas in the previous case, $Z = \frac{q^2}{(1+p^2)^{5/2}}$, now $Z = \frac{Sq^2}{(1+p^2)^{5/2}}$, where $S$ is the *indeterminate magnitude* that is present in the formula $\int \frac{Sds}{R^2}$ of the maximum and minimum. This indeterminate magnitude requires the application of the *differential value of the second kind* (see Chapter IV of the *Methodus inveniendi*).

The *differential value of the second kind* is given by the expression $D$ below, which is found with the aid of expressions $A, B,$ and $C$ as will be shown next.

$$dZ = Ld\Pi + Mdx + Ndy + Pdp + Qdq, \qquad (A)$$

where $\Pi = \int[Z]dx$.

Key to find *the differential of the second kind* is the identification of the quantity $Ld\Pi$, which Euler gives as $Ld\Pi = \frac{q^2 Tds}{(1+p^2)^{5/2}}$, from which $L = \frac{q^2 T}{(1+p^2)^{5/2}}$, and $d\Pi = ds = dx\sqrt{1+p^2}$.

Since $d\Pi = [Z]dx$, it follows that $[Z] = \sqrt{1+p^2}$.

Knowing that

$$d[Z] = [M]dx + [N]dy + [P]dp, \qquad (B)$$

then, $[M] = 0$, $[N] = 0$, $[P] = \frac{p}{\sqrt{1+p^2}}$ and $[Q] = 0$.

Now since $Z = \frac{Sq^2}{(1+p^2)^{5/2}}$, then from $(A)$, we also have that

$$dZ = \frac{q^2 dS}{(1+p^2)^{5/2}} + Pdp + Qdq,$$

where $P = \frac{-5Sq^2 p}{(1+p^2)^{7/2}}$, and $Q = \frac{2Sq}{(1+p^2)^{5/2}}$.

Now the integral may be taken

$$\int Ldx = \int \frac{q^2 Tdx}{(1+p^2)^{\frac{5}{2}}} = \int \frac{q^2 dS}{(1+p^2)^3}$$

and its value shall be $H$ and, since

$$V = H - \int Ldx, \qquad (C)$$

therefore,

$$V = H - \int \frac{q^2 dS}{(1+p^2)^3}.$$

The *differential value of the second kind* is given by

$$-d.(P + [P]V) + \frac{d(Q + [Q]V)}{dx}, \qquad (D)$$

which, in this case, reduces to

$$-dP - d.[P]V + \frac{dQ}{dx} \qquad (II).$$

On account that the two *differential values* $(I)$ and $(II)$ found above should be equal, this equation arises for the curve sought

$$d.\frac{p}{\sqrt{1+p^2}} = dP + d.[P]V - \frac{dQ}{dx},$$

which once integrated gives

$$\frac{\alpha p}{\sqrt{1+p^2}} + \beta = P + [P]V - \frac{dQ}{dx},$$

or

$$\frac{\alpha p}{\sqrt{1+p^2}} + \beta = \frac{Hp}{\sqrt{1+p^2}} - \frac{p}{\sqrt{1+p^2}} \int \frac{q^2 dS}{(1+p^2)^3} + P - \frac{dQ}{dx}.$$

On account that the constant $H$ can be taken to be determined by the arbitrary constant $\alpha$, Euler obtains the differential equation for the curve sought as

$$\frac{\alpha p}{\sqrt{1+p^2}} + \beta = P - \frac{dQ}{dx} - \frac{p}{\sqrt{1+p^2}} \int \frac{q^2 dS}{(1+p^2)^3}.$$

Subsequent integrations and further mathematical manipulations of this equation, finally results in

$$\frac{S}{R} = \alpha + \beta x - \gamma y.$$

Euler then indicates that the quantity $\alpha + \beta x - \gamma y$ expresses the moment of the curving force, which must be equal to the flexural rigidity $S$, divided by the radius of osculation $R$. He then points out having fully vindicated Bernoulli's result – referring to the *Euler-Bernoulli formula* $\mathcal{M} = \frac{EI}{r}$ – and recognizing that "… the more difficult use of my formulas taken in this example has been made clear …"

**Investigations by direct methods**

Inclusions of investigations by direct methods in a variational treatise may seem odd, however, Fraser [9] argues that this was an opportunity for Euler to publish these supplemental researches which were aimed at the theory of elasticity.

Indeed, linked to the results presented in the last section, in § 44 of *Additamentum I* Euler presents an investigation by direct methods of a lamina with variable elasticity, to find the flexural rigidity $S$ at some place $M$ of an unequal elasticity lamina which is curved by the weight $P$ hanging down at the free extremity of the lamina. He then found that that $S = PRx$, where $x$ is the lever arm that extends horizontally from the extremity where the weight is hanging, to the vertical line projected down from $M$. This result is, indeed, confirmed by the variational method presented in the last section.

Next, in § 45, by considering that the absolute elasticity is proportional to the width of the lamina, Euler then examines a triangular shape lamina like a tongue when a weight hangs down at the free extremity but was not able to find a general workable expression for the curvature.

**Vibrations of elastic laminas**

This is another investigation by direct methods considered by Euler in the second part of *Additamentum I* in the section 'Concerning the oscillations of elastic laminas with the other end fixed to a wall'.

This development begins in § 65, where, by equating the harmonic force acting on the lamina to the elastic restoring force, Euler finds a fourth order differential equation governing the vibratory motion of the lamina given by

$$EI \frac{d^4 y}{dx^4} = \frac{M}{af} y,$$

where $M$ is the weight of the elastic band, $a$ its length, and $f$ is the length of the simple isochronous pendulum. This is how Euler introduces the frequency of oscillation into the modeling.

In a more modern notation, this equation may be written as

$$EI\frac{d^4y}{dx^4} = \mu\omega^2 y,$$

where $\mu = \frac{M}{ga}$, is the mass per unit of length, $g$ is the gravity, and $\omega = \sqrt{\frac{g}{f}}$ is the angular frequency.

From this equation, Euler obtains an expression for the spatial solutions (mode shapes) of the vibrating band as

$$y = A\sinh\frac{x}{c} + B\cosh\frac{x}{c} + C\sin\frac{x}{c} + D\cos\frac{x}{c}, \text{[m]}$$

where $A, B, C, D$ are constants of integration which are determined from the boundary conditions. In this expression $c = \left(\frac{EI}{\mu\omega^2}\right)^{1/4}$ has units of length and $c^{-1}$ is the wave number $\beta$.

By applying the proper boundary conditions for the cantilever beam yields the following transcendental equation. There are multiples roots which satisfy this equation. A subscript is thus added.

$$\cos(\beta_n a)\cosh(\beta_n a) = -1. \text{[n]}$$

Euler obtains the first four roots of this equation as $\beta_1 a = 1.8751040818$, $\beta_2 a = 4.6940910795$, $\beta_3 a = 7,8547670321$, $\beta_4 a = 10,9955428716$.

The corresponding natural frequencies are

$$\omega_1 = \beta_1^2\sqrt{\frac{EI}{\mu}} = \frac{3.5160}{a^2}\sqrt{\frac{EI}{\mu}}, \ldots$$

4. **Final Comments**

Levien [2] demonstrates the pervasiveness of the elastica solution through various examples.

- It appears in another shape of the solution of a fundamental physics problem – the capillary. In 1807, Laplace had obtained an equation equivalent to Euler's result $\frac{S}{R} = \alpha + \beta x - \gamma y$ deduced above, for the curvature of the surface of a fluid trapped between two vertical plates.
- The differential equation for the elastica, expressing curvature as a function of arclength, is equivalent to those of the motion of the pendulum, as worked out by Kirchhoff in 1859 – the differential equation for the shape of the elastica is mathematically equivalent to that of the dynamics of a simple swinging pendulum.

---

[m] In fact, Euler gives a less general equation as $y = Ae^{\frac{x}{c}} + Be^{-\frac{x}{c}} + C\sin\frac{x}{c} + D\cos\frac{x}{c}$.

[n] In fact, Euler replaces this equation by $e^{(\beta_n a)} = \frac{-1 \pm \sin(\beta_n a)}{\cos(\beta_n a)}$.

- Despite the equation for the general elastica being published as early as 1695, the curves had not been accurately plotted until Max Born's 1906 PhD thesis. Born also constructed an experimental apparatus using weights and dials to place the elastic band in different endpoint conditions and used photographs to compare the equations to actual shapes.

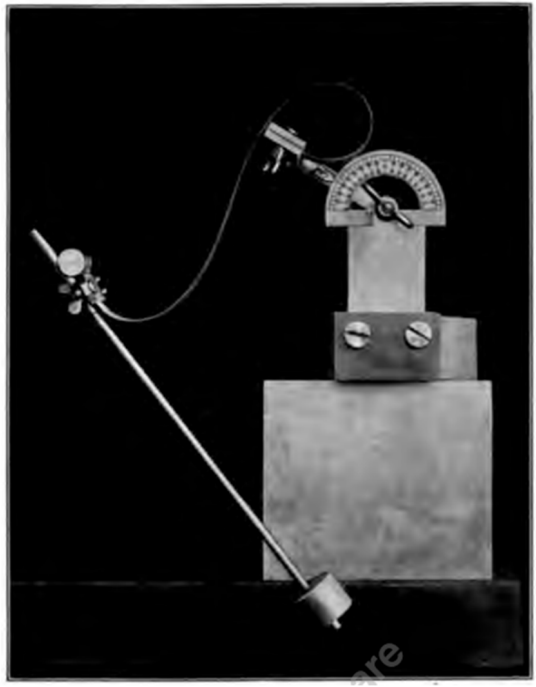

Figure 6: Born's experimental apparatus for measuring the elastica. (Source: Ref. [2])

- Influence on modern spline theory: mechanical splines made of wood or metal have long been an inspiration for the mathematical concept of spline (and for its name). Schoenberg's justification for cubic splines in 1946 was a direct appeal to the notion of an elastic strip. His main contribution was to define his spline in terms of a variational problem closely approximating the minimization of potential energy adopted in the elastica solution.
- The arrival of the high-speed digital computer created a strong demand for efficient algorithms to compute the elastica, particularly to compute the shape of an idealized spline constrained to pass through a sequence of control points.

As Levien [2] points out, "… the elastica, having been present at the birth of the variational calculus, also played a major role in the development of another branch of mathematics: the theory of elliptic functions. Even as the quadratures of these simple curves came to be revealed, analytic formulae for their lengths remained elusive. The functions known by the first half of the 18th century were insufficient to determine the length even of a curve as well-understood as an ellipse …" The closed-form solutions of the elastica, worked out in the 1880s, rely heavily on Jacobi elliptic functions. Nonetheless, elliptic functions are rarely used today for computation of elastica, in favor of numerical methods. To this end, simple 4th-order Runge-Kutta differential equation solver can be used due to its good convergence and efficiency and simple expression in code. "… Even so, Jacobi elliptic functions are now part of the mainstream of special functions, and fast algorithms for computing them are well-known. Elliptic functions are still the method of choice for the fastest computation of the shape of an elastica …"

Finally, as revealed by D'Antonio [3], Euler's work on elastic curves has been extended to study the elastic properties of the DNA molecule by considering it as a symmetric elastic rod. The unstressed DNA molecule has the form of a double helix with a twist angle of approximately 34º. If one end of the DNA molecule is twisted, an elastic strain is induced which tends to untwist it to the equilibrium configuration, e.g., a figure of eight.

Note: this historical sketch is based on primary sources (and their annotated English translations), but also draws heavily on secondary sources, all listed in the References.

**Declarations**

No funding was received for conducting this study.

The author has no competing interests to declare that are relevant to the content of this article.